\documentclass{desyproc}

\usepackage{subfigure}

\newcommand{\lsim}{\lesssim}
\newcommand{\gsim}{\gtrsim}

\newcommand{\ord}[1]{\mathcal{O}{(#1)}}
\newcommand{\beq}{\begin{equation}}
\newcommand{\eeq}{\end{equation}}
\newcommand{\eps}{\varepsilon}

\newcommand{\ald}{\alpha_d}

\begin{document}
\title{Implications of a Running Dark Photon Coupling}

\author{{\slshape  Hooman Davoudiasl$^{1}$}\\[1ex]
$^1$ Department of Physics, Brookhaven National Laboratory, Upton, NY 11973, USA
}

\contribID{familyname\_firstname}

\confID{11832}  
\desyproc{DESY-PROC-2015-02}
\acronym{Patras 2015} 
\doi  

\maketitle

\begin{abstract}

For an ``invisible" dark photon $Z_d$ that dominantly decays into dark states, the running of its
fine structure constant $\ald$ with momentum transfer $q > m_{Z_d}$ could be significant.  A 
similar running in the kinetic mixing parameter $\eps^2$ can be induced through its dependence on $\ald(q)$.  The 
running of couplings could potentially be detected in ``dark matter beam" experiments, for which    
theoretical considerations imply $\ald (m_{Z_d}) \lsim 0.5$.

\end{abstract}


The following is a summary of a talk - entitled ``Running in the Dark Sector" - given by the author at the 11$^{\rm th}$ Patras Workshop on 
Axions, WIMPs and WISPs, held at the University of Zaragoza, Spain, June 22-26, 2015.  The 
presentation is based on the work in Ref.\cite{Davoudiasl:2015hxa}, where a more complete 
set of references can be found.   

The possibility of a dark sector that includes not only dark matter (DM), but also dark forces and other 
states has attracted a great deal of attention in recent years \cite{Essig:2013lka}.  In particular, it has been noted that a ``dark photon" $Z_d$ 
of mass $m_{Z_d} \lsim 1$~GeV, mediating a dark sector $U(1)_d$ force
may explain potential astrophysical signals of DM \cite{ArkaniHamed:2008qn}.   
It is often assumed that the $Z_d$ can couple to the 
electromagnetic current of the Standard Model (SM) via a small amount of kinetic mixing 
$\eps$ \cite{Holdom:1985ag} (though it may have other couplings as well \cite{Davoudiasl:2012ag}) 
which can be naturally loop induced: $\eps \sim e g_d/(16 \pi^2)$ \cite{Holdom:1985ag} where $e$ and 
$g_d$ are the electromagnetic and $U(1)_d$ and coupling constants, respectively.  
The 3.5\,$\sigma$ muon $g-2$ anomaly \cite{Bennett:2006fi} may potentially 
be explained by a light ($m_{Z_d} \lsim 0.1$~GeV) $Z_d$ with $\eps\sim 10^{-3}$ \cite{Pospelov:2008zw}.  

\begin{figure}[ht]
\centerline
{\includegraphics[width=0.8\textwidth]{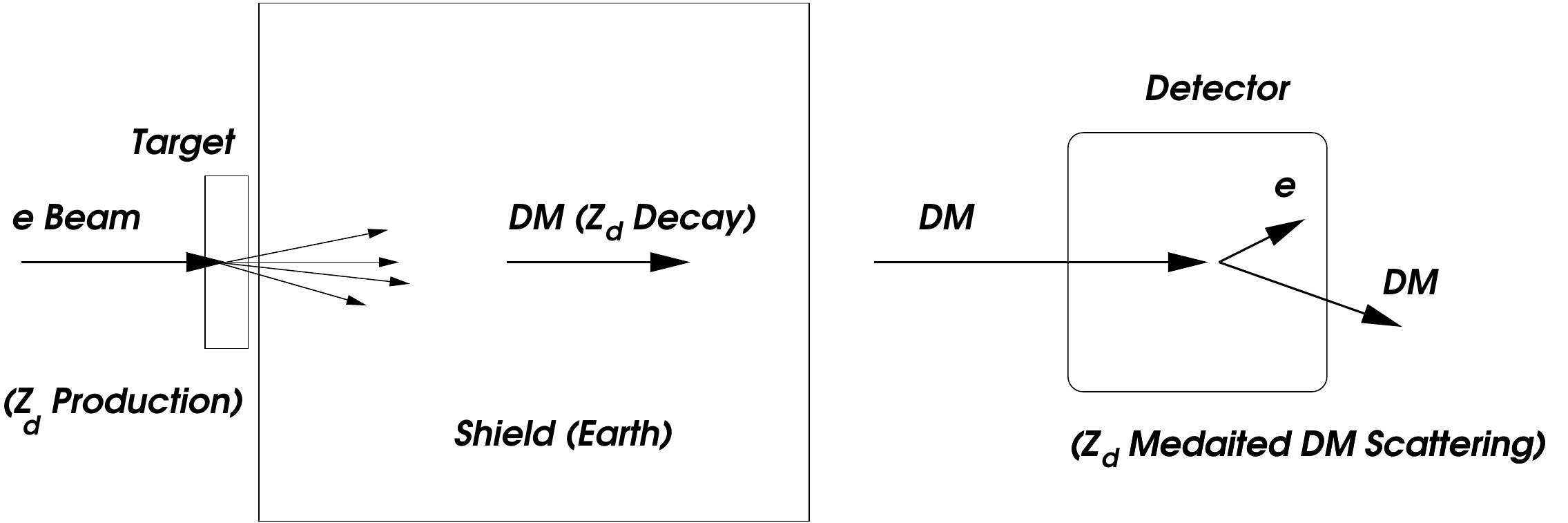}}
\caption{Schematic illustration of a fixed target ``dark matter beam" experiment, using an electron beam.}
\label{darkbeam}
\end{figure}

If there are dark states, such as DM, that have $U(1)_d$ charge $Q_d\neq 0$ and have a mass $m_d < m_{Z_d}/2$, then 
they will likely be the dominant decay channels for $Z_d$, making it basically invisible.  This possibility can be employed to 
form beams of light (sub-GeV) DM that may be detectable in fixed target experiments 
(whose detection in nuclear recoil experiments would be challenging).  The basic idea is that an intense 
beam of protons or electrons impinging on a target (or beam dump) can lead to production of boosted $Z_d$ particles  
that decay in flight mostly into light DM states, generating a ``DM beam" which can be detected via $Z_d$-mediated 
scattering from atoms \cite{deNiverville:2011it,Izaguirre:2014bca}.  See Figure \ref{darkbeam} for a schematic illustration of 
such a setup.  The production rate of on-shell dark photons is controlled by $\alpha \eps^2$, while the detection of the DM particles 
is governed by $\ald\alpha\eps^2$, where $\alpha\equiv e^2/(4\pi)$ and $\ald\equiv g_d^2/(4 \pi)$.  
\begin{figure*}[ht]
\begin{center}
\subfigure[]{
\includegraphics[width=0.45\textwidth,clip]{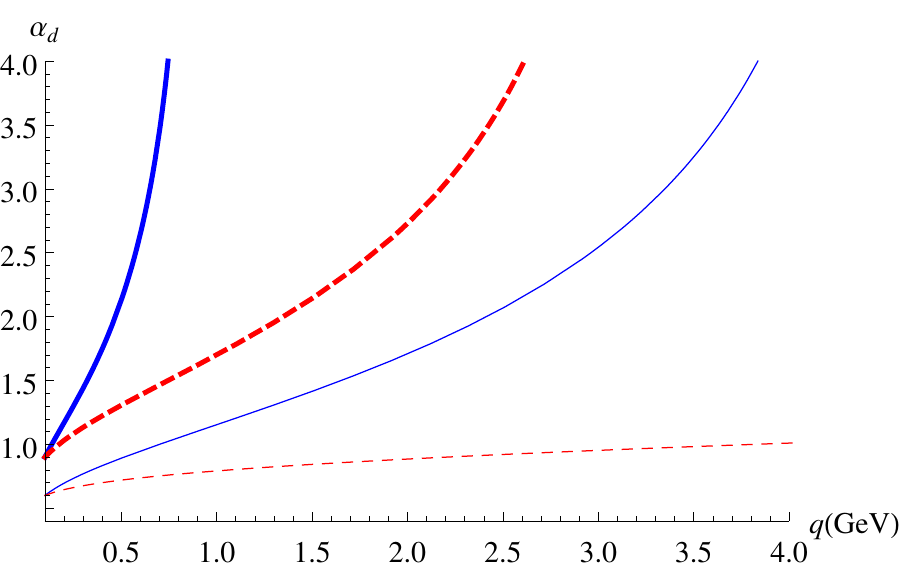}}
~~~~~
\subfigure[]{
\includegraphics[width=0.45\textwidth,clip]{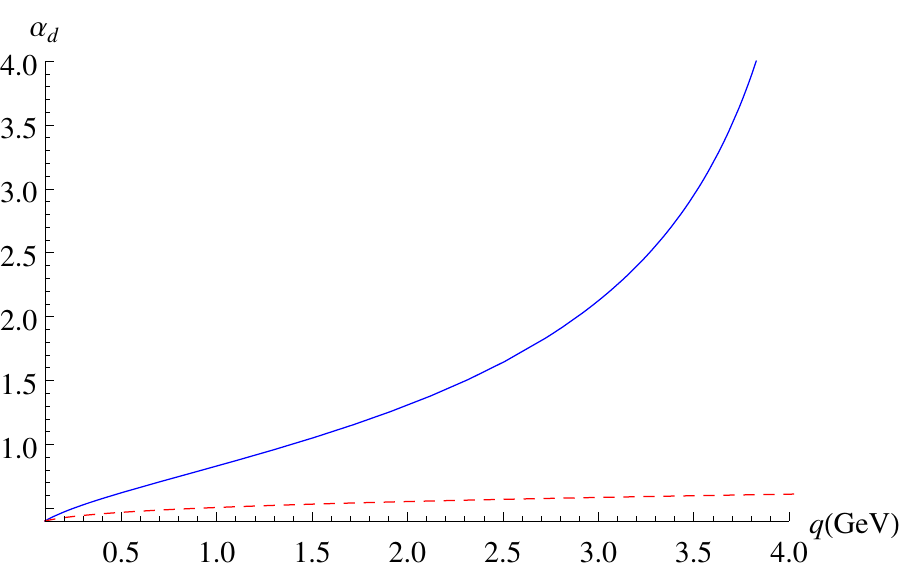}}
\end{center}
\caption{Running of $\ald(q)$, with (a) one DM particle, where the thin (thick) lines correspond to
$\ald(q_0) = 0.6$ (0.9), and (b) two DM states with $\ald(q_0)=0.4$.  The solid (dashed)
lines correspond to fermion (scalar) DM states.  
In both cases, the contribution from a dark Higgs particle is included, 
$q_0=0.1$~GeV, and $m_{Z_d}\lsim q_0$ is assumed.  }
\label{ald-run}
\end{figure*}
\begin{figure}[ht]
\centerline{
\includegraphics[width=0.45\textwidth]{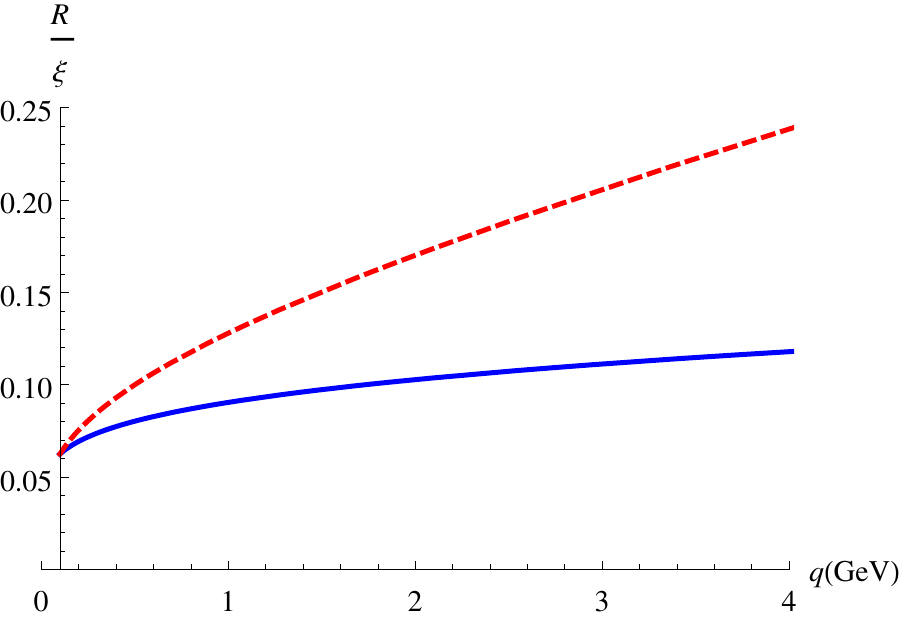}}
\caption{Running of $R/\xi$ with $q$, assuming one (solid) and two (dashed) 
dark matter fermions.  Here, $\ald(q_0) = 0.25$, $q_0=0.1$~GeV, and $m_{Z_d}\lsim q_0$ are assumed.
A dark Higgs boson contributes to the running in both cases.}
\label{roverxi}
\end{figure}

If the above light DM particles are thermal relics, one expects \cite{deNiverville:2011it,Izaguirre:2014bca}
\beq
\ald \sim 0.02\, w \left(\frac{10^{-3}}{\eps}\right)^2 \left(\frac{m_{Z_d}}{100~\text{MeV}}\right)^4
\left(\frac{10~\text{MeV}}{m_d}\right)^2\,,
\label{ald}
\eeq
where $w\sim 10$ for a complex scalar \cite{deNiverville:2011it}, and $w\sim 1$ for a fermion \cite{Izaguirre:2014bca}. 
As experiments probe smaller values of $\eps$, one could start probing $\ald \sim 1$, which in the presence of light DM with 
$Q_d\neq 0$ can lead to significant running of $\ald$.  As the fixed target experiments (Figure \ref{darkbeam}) probe 
momentum transfer values in the GeV range, {\it i.e.} $q^2 \gg m_{Z_d}^2$, the effect of running 
on the event rate can be significant and it may even 
lead to unreliable predictions for $\ald(q^2)\gg 1$.  To illustrate these points, we will consider $n_F$ 
fermions and $n_S$ scalars with $|Q_d|=1$, all below $m_{Z_d}$.  We will assume that the mass of $Z_d$ is generated by a dark Higgs 
scalar and hence $n_S\geq 1$ in our analysis.

We will employ a 2-loop beta function for $U(1)_d$
\beq
\beta(\ald) = \frac{\ald^2}{2\pi}\left[\frac{4}{3}\left(n_F + \frac{n_S}{4}\right) +
\frac{\ald}{\pi} (n_F + n_S)\right]\,,
\label{beta}
\eeq
where $\beta(\ald)\equiv \mu \, d\ald/d\mu$, with $\mu$, the renormalization
scale, set by the relevant momentum transfer $q$.   The reference infrared  
momentum transfer is taken to satisfy $q_0 \gsim m_{Z_d}$ and will ignore the mass of $Z_d$ in what follows.   
The form of $\beta(\ald)$ in the above suggests that perturbative control is lost when $\ald\gsim \pi$.   

In Figures \ref{ald-run} (a) and (b), we have presented the effect of running for various 
values of $\ald$ and one and two DM states, respectively.  We see that for values of $\ald \lsim 1$ 
the running effect can be significant and may result in loss of perturbative reliability for predictions.  The running is 
more pronounced for light fermion states, but could still be significant for scalars for $\ald\gsim 0.4$.  
These results suggest that one may be able to use the running effect, if measurable, to probe the number and 
the type (spin) of the low lying states in the dark sector.

An approach to measuring the running of $\ald(q)$ may take advantage of the fact that 
at $q^2 > m_{Z_d}$ the scattering of DM from the nucleus is similar 
to electron or muon electromagnetic scattering from the nucleus 
governed by quantum electrodynamics.  One may then normalize the DM 
scattering cross section $\sigma_{\rm DM}$ to the well-understood 
lepton scattering cross section $\sigma_{\rm EM} \propto 1/q^2$ which can be well-measured.  
We then have 
\beq
R\equiv \sigma_{\rm DM}/\sigma_{\rm EM} \simeq \ald\, \eps^2/\alpha
\simeq \xi\, \ald^2\, ,
\label{R}
\eeq
with $\xi$ approximately constant. In the above, we have used the typical assumption of loop-induced kinetic mixing   
that implies $\eps^2 (q)\propto \ald (q)$.  In Figure \ref{roverxi}, we have plotted the running of 
$R/\xi$ for one (solid) and two (dashed) light   
DM fermions and one dark Higgs boson, assuming 
$\ald (q_0) = 0.25$, $q_0=0.1$~GeV, and $m_{Z_d} \lsim q_0$.  As can be seen, the running is significant 
for GeV~$0.1 \lsim q \lsim 4$~GeV, typical of fixed target experiments, and the 
two cases are quite distinct, suggesting that with sufficient statistics one may 
uncover the low lying dark sector spectrum.  

As $\ald$ increases beyond $\ord{1}$ values, the theory will become strongly coupled.  However, in a sensible 
framework, this behavior should be terminated at some scale.  A straightforward possibility is 
for $U(1)_d$ to transition to a non-Abelian gauge interaction that is asymptotically free.  If this transition to new physics 
occurs at $q=q^*$, one expects $\eps (q^*) = 0$, with a non-zero value induced below $q^*$ 
due to the quantum effects of particles with masses $m<q^*$ that carry hypercharge and have $Q_d\neq 0$.  
However, such particles cannot be too light, $m \gsim 100$~GeV \cite{Davoudiasl:2012ig}, given existing experimental bounds.  
Thus, on general grounds, we expect $q^*$ to be larger than $\ord{100\, \text{GeV}}$.  

For $\ald(q^*) \ln(q^*/q_0)\gg 1$, we find
\beq
\ald (q_0) \approx \frac{3\pi}{(2 n_F + n_S/2)\ln(q^*/q_0)}\,,
\label{aldqest}
\eeq 
where we have used a 1-loop approximation for the running.  The above formula then yields 
the value of $\ald(q_0)$ that would lead to the onset of a Landau pole at $q\sim q^*$.  For example, 
setting $q_0=0.1$~GeV and $q^* = 1$~TeV (a reasonable value given the preceding discussion), the  
upper bound $\ald(q_0) \lsim 0.5/(n_F + n_S/4)$ is obtained.  Hence, for $m_{Z_d} \lsim 0.1$~GeV, we 
may expect the upper bound $\ald(m_{Z_d}) \lsim 0.5$ as a generic guide for the invisible 
dark photon scenario, where dark states below $m_{Z_d}$ are assumed.

\section*{Acknowledgments}

Th author thanks the organizers of Patras 2015 for giving him the opportunity to present the above results 
and for providing a pleasant venue for stimulating discussions.  This article is based on work supported by the US Department of 
Energy under Grant Contract DE-SC0012704.



\begin{footnotesize}

\end{footnotesize}


\end{document}